\documentclass[preprint,nofootinbib,longbibliography,amsmath,amssymb,aps,prd]{revtex4-2}

\usepackage{comment}
\usepackage[colorlinks,linkcolor=blue,citecolor=blue]{hyperref}
\usepackage{amsmath,amssymb,amsfonts,graphicx,fancyhdr,epstopdf}
\usepackage[top=2cm,bottom=2cm,left=2cm,right=2cm]{geometry}
\usepackage{multirow}
\usepackage{hhline}
\usepackage{threeparttable}
\usepackage{mathrsfs}
\usepackage{dsfont}
\usepackage{cancel} 
\usepackage{ulem}
\usepackage{romannum}
\usepackage{subfigure}
\usepackage{adjustbox}
\usepackage{dcolumn}
\usepackage{bm}
\usepackage{enumerate}

\usepackage{xcolor}
\usepackage{float}
\usepackage{enumitem}  
\usepackage{bbold}
\usepackage{tikz}
\usepackage{supertabular,rotating,adjustbox}
\newcommand{\revs}[1]{{\color{blue}#1}}

\begin{document}

	\preprint{APS/PRD}
	\title{\boldmath
		The Benchmark Mode  $\Omega_c \to \Omega^-\pi^+$ and  Its Related Processes}

	\author{Shuge Zeng}
	\author{Fanrong Xu}
	\email{Electronic address: fanrongxu@jnu.edu.cn}
	
         \author{Yu Gu}

	\affiliation{Department of Physics, College of Physics $\&$ Optoelectronic Engineering, 
	Jinan University, Guangzhou 510632, P.R. China}

	\begin{abstract} 
 
 The benchmark mode  $\Omega_c^0\to \Omega^- \pi^+$, which receives purely factorization contribution, 
 is of great importance among all the decay channels of $\Omega_c^0$ decays. In this work,
 within the framework of non-relativistic quark model (NRQM),
 we calculate all the 6 baryon transition form factors involving
 $\frac12 ^+\to \frac32 ^+$  decays.  
 The absolute branching fractions of non-leptonic decays $\Omega_c^0\to \Omega^- \pi^+$, $\Omega_c^0\to \Omega^- \rho^+$ and
 $\Omega_c^0\to \Xi^- \pi^+$ as well as semi-leptonic decays $\Omega_c^0\to \Omega^- \ell^+ \nu_{\ell} \; (\ell=e,\mu)$ are calculated
 although they cannot be measured directly by current experiment.
Based on the prediction  $\mathcal{B}(\Omega_c^0\to\Omega^-\pi^+)=(3.43\pm0.48)\%$ in our work,
we further predict the ratios between interested modes and  the benchmark mode, 
giving $R(\Xi^-\pi^+)=0.16\pm0.06$, $R(\Omega^-\rho^+)=5.33\pm0.94$, $R(\Omega^- e^+\nu_e)=1.18\pm0.22$ and
$R(\Omega^- \mu^+\nu_\mu)=1.11\pm0.20$. 
 The predictions on $\Omega_c^0\to \Xi^-\pi^+$ 
 and $\Omega_c^0\to \Omega^- e^+\nu$ agree well
 with recent measured ratios reported by LHCb in 2023 and ALICE in 2024, respectively.
	
	\end{abstract}
	
	\maketitle
	\clearpage
	\newpage
	\pagenumbering{arabic}

\section{Introduction}
\label{sec:intro}


Significant experimental progress has been made in the study of charmed baryon decays in recent years. For the anti-triplet singly charmed baryons $\Lambda_c$ and $\Xi_c^{0,-}$, substantial experimental data has been accumulated since the measurements of their benchmark decay modes 
$\Lambda_c\to p K^- \pi^+$ \cite{Belle:2013jfq} \cite{BESIII:2015bjk} and $\Xi_c^0\to \Xi^- \pi^+$ \cite{Belle:2018kzz}.
As the lightest member of the singly charmed baryon sextet, the $\Omega_c$ baryon primarily decays via weak processes. 
Although its benchmark mode $\Omega_c^0\to \Omega^- \pi^+$ has not yet been measured, 
a relative ratio to this benchmark mode, defined as
\begin{equation}
R(X)=\frac{\mathcal{B}(\Omega_c^0 \to X)}
{\mathcal{B}(\Omega_c^0 \to \Omega^-\pi^+)},
\end{equation}
is experimentally less challenging to determine. 
By measuring $R(X)$, information related to $\Omega_c$ can still be indirectly extracted.
Recent measurements have provided results such as:
\begin{equation}
\begin{split}
{\rm{LHCb}}\, 2023 ~\text{\cite{LHCb:2023fvd}}:\quad  &
 10^2R(\Omega^- K^+) = 6.08\pm 0.51\pm 0.40,\; \;  \\
& 10^2R(\Xi^- \pi^+)=15.81\pm0.87\pm 0.43\pm 0.16 \\
{\rm{ALICE}}\, 2024 ~\text{\cite{ALICE:2024xjt}}:\quad 
& 10^2R(\Omega^- e^+ \nu_e)=1.12\pm0.22\pm0.27.
\end{split}
\end{equation}
Incorporating previously measured results, Table \ref{tab:exp} summarizes all the experimental progress on $\Omega_c$ decays to date.

The recent experimental progress necessitates concurrent 
theoretical studies on $\Omega_c$ decays. To align with the current experimental status, estimating the branching fraction of the benchmark mode is both timely and essential, taking priority over calculating the absolute branching fractions of other modes of interest. The mode $\Omega_c^0 \to \Omega^- \pi^+$ is particularly noteworthy due to its kinematics and dynamics. Classified as a $\frac{1}{2}^+ \to \frac{3}{2}^+ + 0^-$ decay, this benchmark mode provides rich kinematic information. Furthermore, it receives contributions only from purely factorizable processes, meaning non-perturbative effects are solely described by baryon transition form factors. 

Theoretical studies on $\Omega_c$ decays date back to the 1990s \cite{Xu:1992sw, Cheng:1996cs}. However, there was a lull in theoretical efforts until recent years, when renewed interest was sparked by rapid experimental developments. For recent theoretical studies of  $\Omega_c$, 
various phenomenology methodologies have been applied,
 including the light-front quark model \cite{Hsiao:2020gtc, Zhao:2018zcb}, constitute quark model \cite{Wang:2022zja}, light-cone sum rules \cite{Aliev:2022gxi}, and fit in combination with other techniques \cite{Liu:2023dvg, Hsiao:2023mvw}.
In our previous series of work \cite{Cheng:2018hwl,Zou:2019kzq,Cheng:2020wmk,Meng:2020euv,Cheng:2022kea}, 
within the framework of the topological diagrammatic assisted pole model, we systematically calculated  
absolute branching fractions of all $\frac{1}{2}^+ \to \frac{1}{2}^+ + 0^-$ decays  using current algebra combined with the MIT bag model,
including decays of $\Omega_c$ \cite{Hu:2020nkg}. However, due to the absence of a benchmark mode calculation, our predictions 
related to $\Omega_c$ could not be directly compared with current experimental measurements.
In this work, we aim to bridge this gap by incorporating the reference mode and other related channels within the non-relativistic quark model (NRQM). 
The choice of NRQM is particularly reasonable for studying charmed baryon decays 
due to the relatively low momentum carried by the decay products.  
A previous theoretical calculation within NRQM for  doubly charmed baryon decays also
provided predictions in agreement well with experimental results \cite{Zeng:2022egh}.
Therefore, by consistently applying NRQM to all related modes of $\Omega_c$ decays, 
the reliability  of our predictions are expected to be enhanced.

\begin{table}[t!]
\centering
\caption{ A summary of  measured $R(X)$ in recent experiments.} 
\label{tab:exp}
\vspace{1pt}
\setlength{\tabcolsep}{0.1mm}
\scriptsize{
\begin{tabular}{|cl|clclclclclcl|}
\hline\hline
\multicolumn{2}{c}{$\qquad$Results$\qquad$} & \multicolumn{2}{c}{ALICE(2024)\cite{ALICE:2024xjt}}& \multicolumn{2}{c}{$\;$LHCb(2023)\cite{LHCb:2023fvd}$\;$} & \multicolumn{2}{c}{Belle(2022)\cite{Belle:2022yaq}} &  \multicolumn{2}{c}{Belle(2021)\cite{Belle:2021dgc}} &\multicolumn{2}{c}{Belle(2017)\cite{Belle:2017szm, PDG:2024}}& \multicolumn{2}{c}{$\quad$CLEO(2002)\cite{CLEO:2002imi}$\quad$}\\ \hline
\multicolumn{2}{c}{$R(\Omega^-e^+\nu_e)$}& \multicolumn{2}{c}{$1.12\pm 0.22\pm 0.27$}  & \multicolumn{2}{c}{}   & \multicolumn{2}{c}{}& \multicolumn{2}{c}{$1.98\pm 0.13\pm 0.08$}& \multicolumn{2}{c}{}& \multicolumn{2}{c}{$2.439\pm1.154$} \\ 
\multicolumn{2}{c}{$R(\Omega^- \mu^+ \nu_\mu)$} & \multicolumn{2}{c}{}  & \multicolumn{2}{c}{}&\multicolumn{2}{c}{}& \multicolumn{2}{c}{$1.94\pm 0.18\pm 0.10$}& \multicolumn{2}{c}{}& \multicolumn{2}{c}{}\\ \hline
\multicolumn{2}{c}{$10^2R(\Xi^-\pi^+)$} & \multicolumn{2}{c}{}& \multicolumn{2}{c}{$15.81\pm 0.97\pm 0.16\quad$}& \multicolumn{2}{c}{$25.3\pm 5.2 \pm 3.0$}    & \multicolumn{2}{c}{}& \multicolumn{2}{c}{}& \multicolumn{2}{c}{}\\
\multicolumn{2}{c}{$10^2R(\Omega^-K^+)$} & \multicolumn{2}{c}{} & \multicolumn{2}{c}{$6.08\pm 0.51\pm 0.40$} & \multicolumn{2}{c}{$<29$}  &\multicolumn{2}{c}{}& \multicolumn{2}{c}{}& \multicolumn{2}{c}{}\\
\multicolumn{2}{c}{$10^2 R(\Xi^-K^+)$} & \multicolumn{2}{c}{}  & \multicolumn{2}{c}{}  & \multicolumn{2}{c}{$<7$} & \multicolumn{2}{c}{}& \multicolumn{2}{c}{}& \multicolumn{2}{c}{}\\ 
\multicolumn{2}{c}{$R(\Omega^-\rho^+)$} & \multicolumn{2}{c}{}  & \multicolumn{2}{c}{}  & \multicolumn{2}{c}{}& \multicolumn{2}{c}{}& \multicolumn{2}{c}{$>1.3$}& \multicolumn{2}{c}{} \\ \hline\hline
\end{tabular}
}
\end{table}

The structure of this paper is organized as follows: Section \ref{sec:kin} begins with a revision of the kinematics associated with non-leptonic and semi-leptonic decays involving $\frac{1}{2}^+ \to \frac{3}{2}^+$. In Section \ref{sec:FF}, we derive the analytical expressions for form factors within the non-relativistic constituent quark model. Section \ref{sec:num} presents numerical analyses of these form factors, calculates the branching fractions for the benchmark mode, and explores other related non-leptonic and semi-leptonic decays, including their relative ratios. Conclusions are drawn in Section \ref{sec:con}. The appendices provide additional support. Appendix \ref{app:FFCon} discusses different conventions and their interrelations, while Appendix \ref{app:FFcalc} details the calculation of form factors within the non-relativistic quark model.

\section{Kinematics}
\label{sec:kin}


The benchmark mode $\Omega_c^0 \to \Omega^- \pi^+$ is categorized as a $\frac{1}{2}^+ \to \frac{3}{2}^+ + 0^-$ process, with particular focus on the decuplet baryon $\frac{3}{2}^+$. In this section, we first derive the general kinematics for two-body non-leptonic decays, covering final states that include both pseudoscalar and vector mesons. Additionally, we provide generic kinematic formulas for semi-leptonic decays.

\subsection{Non-leptonic decays}
\label{subsec:nl}

The effective Hamiltonian for Cabibbo-favored (CF) processes in $\Omega_c^0$ decay is given by:
\begin{align}
\mathcal{H}_{\text{eff}} = \frac{G_F}{\sqrt{2}} V_{cs} V_{ud}^\ast (c_1 O_1 + c_2 O_2) + \text{H.c.}
\end{align}
with the four-quark operators:
\begin{align}
O_1 = (\bar{s}c)(\bar{u}d) \;, \qquad O_2 = (\bar{s}d)(\bar{u}c) \;,
\end{align}
where $(\bar{q}_1 q_2) = \bar{q}_1 \gamma_\mu (1-\gamma_5) q_2$. The leading-order Wilson coefficients are $c_1 = 1.346$ and $c_2 = -0.636$~\cite{Cheng:2018hwl}. For convenience, the effective Wilson coefficients $a_1 = c_1 + c_2/N_{\rm{eff}}$ and $a_2 = c_2 + c_1/N_{\rm{eff}}$ are introduced.

Without delving into the specifics of the dynamics, the decay amplitudes for $\frac{1}{2}^+ \to \frac{3}{2}^+$ transitions can be parameterized generically as:
\begin{equation}
\begin{split}
& M (\mathcal{B}_i \to \mathcal{B}_f P) = i q_{\mu} \bar{u}_f^{\mu}(p_f)(C+D\gamma_5)u_i(p_i) \;, \\
& M (\mathcal{B}_i \to \mathcal{B}_f V) = \bar{u}_f^{\nu}(p_f) \epsilon^{*\mu} \Big[ g_{\nu\mu}(C_1+D_1\gamma_5) + p_{i\nu}\gamma_\mu(C_2+D_2\gamma_5) + p_{i\nu}p_{f\mu}(C_3+D_3\gamma_5) \Big] u_i(p_i) \;,
\end{split}
\end{equation}
for pseudoscalar and vector mesons in the final state, respectively \cite{Cheng:1996cs}. The final state spin-$\frac{3}{2}$ baryon $\mathcal{B}_f$ is described by the Rarita-Schwinger vector spinor $u^{\mu}$ with momentum $p_f$, while the initial baryon $\mathcal{B}_i$ is characterized by momentum $p_i$. Due to its Lorentz structure, more terms appear in the $\frac{1}{2}^+ \to \frac{3}{2}^+ + 1^+$ transition, hence it is more convenient to use the helicity amplitude framework. 

The decay widths can be expressed as
\begin{equation}
\Gamma(\mathcal{B}_i \to \mathcal{B}_f M) = \frac{p_c}{16\pi m_i^2} H_M, \quad (M=P, V),
\label{eq:width}
\end{equation}
where
\begin{equation}
\begin{split}
& H_P = \frac{Q_+ Q_-}{3 m_f^2} \left( Q_+ |C|^2 + Q_- |D|^2 \right), \\
& H_V = \Bigg\{ 2 \left[ Q_+ \left| C_1 \right|^2 + Q_- \left| D_1 \right|^2 \right] + \frac{2}{3} \left[ Q_+ \left| C_1 - \frac{Q_-}{m_f} C_2 \right|^2 + Q_- \left| D_1 - \frac{Q_+}{m_f} D_2 \right|^2 \right] \\
& \hspace{1cm} + \frac{1}{3 m_f^2 q^2} \Bigg[ Q_+ \left| (M_+ M_- - q^2) C_1 + Q_- M_+ C_2 + \frac{Q_+ Q_-}{2} C_3 \right|^2 \\
& \hspace{2cm} + Q_- \left| (M_+ M_- - q^2) D_1 - Q_+ M_- D_2 + \frac{Q_+ Q_-}{2} D_3 \right|^2 \Bigg] \Bigg\}. \\
\label{eq:HM}
\end{split}
\end{equation}
In Eq.~(\ref{eq:width}), the momentum $p_c$ is given by $p_c = \frac{1}{2m_i} \lambda^{1/2}(m_i^2, m_f^2, q^2)$, where the K\"all\'en function is defined as $\lambda(x,y,z) = x^2 + y^2 + z^2 - 2xy - 2xz - 2yz$. For two-body decays, the on-shell condition $q^2 = m_{P,V}^2$ applies, making $p_c$ a constant determined by the masses of the initial and final baryons. The other related notations in Eq.~(\ref{eq:HM}) are given as $M_{\pm} = m_i \pm m_f$ and $Q_{\pm} = M_{\pm}^2 - q^2$. Evidently, $H_M$ depends on both the baryon and meson in the final state.

\subsection{Semi-leptonic decays}
\label{subsec:sl}

The decay width of semi-leptonic decays can be generally written as
\begin{equation}
\begin{split}
\Gamma(\mathcal{B}_i \to \mathcal{B}_f \ell^+ \nu_\ell) 
&= \frac{1}{192 \pi^3 m_i^2} \int_{m_\ell^2}^{(m_i - m_f)^2} \frac{(q^2 - m_\ell^2)^2 p_c}{q^2} H_\ell \, dq^2, \\
H_\ell &= \left(1 + \frac{m_\ell^2}{2 q^2} \right) b_V H_V + \frac{3 m_\ell^2}{2 q^2} b_P H_P, 
\end{split}
\label{eq:sl}
\end{equation}
where $p_c$ is not a constant due to the off-shell lepton pair momentum transfer $q^2$ in three-body decays. As a key component of the phase space integral, $H_\ell$ can be divided into vector and scalar parts, denoted as $H_V$ and $H_P$ (or $H_M$ in general), with corresponding coefficients
\begin{equation}
b_M = \frac{2}{|V_{ud}|^2 a_1^2 f_M^2 m_M^2}, \qquad (M = P, V).
\end{equation}
It is worth mentioning that the coefficient $b_M$ establishes a connection between hadronic decay and semi-leptonic decay by removing the meson information from $H_M$ defined in Eq.~(\ref{eq:HM}), including the meson decay constant $f_M$ and effective Wilson coefficient $a_1$ (or $a_2$), introduced in Cabibbo-favored non-leptonic decay processes.

\subsection{The parameterization of amplitudes and form factors}
\label{subsec:FF}

It is widely accepted that both factorizable and non-factorizable amplitudes contribute to general charmed baryon decay processes. In the special case where the decay amplitude only receives factorizable contributions, the partial wave amplitudes can be expressed in terms of form factors and the decay constant, giving
\begin{equation}
\begin{split}
C &= -\lambda a_1 f_P \left[\bar{g}_1(m_P^2) + (m_i - m_f) \bar{g}_2(m_P^2) + (m_i E_f - m_f^2) \bar{g}_3(m_P^2)\right], \\
D &= \lambda a_1 f_P \left[\bar{f}_1(m_P^2) - (m_i + m_f) \bar{f}_2(m_P^2) + (m_i E_f - m_f^2) \bar{f}_3(m_P^2)\right], \\
C_i &= -\lambda a_1 f_V m_V \bar{g}_i(m_V^2), \\
D_i &= \lambda a_1 f_V m_V \bar{f}_i(m_V^2),
\end{split}
\label{eq:pwa}
\end{equation}
which depend on the form factors and the  meson decay constant  $f_M  (M=P,V) $ defined as
\begin{equation}
\begin{split}
&\langle 0| \bar{q}_1 \gamma_\mu \gamma_5 q_2 | P(p) \rangle = i p_\mu f_P\;,\\
&\langle 0| \bar{q}_1 \gamma_\mu q_2 | V(\lambda) \rangle = m_Vf_V\epsilon_{\mu}^\lambda\;.
\end{split}
\end{equation}
For the focused $\Omega_c \to \Omega^-$ process in this work, the effective Wilson coefficient is \( a_1 \) while the CKM factor is defined as \( \lambda = \frac{G_F}{\sqrt{2}} V_{cs} V_{ud}^* \), together with the final baryon energy $ E_f = \frac{m_i^2 + m_f^2 - m_M^2}{2m_i} $. There are several different conventions for related baryon transition form factors, although physical observables are independent of form factor parameterization. In this work, the form factors are parameterized as
\begin{equation}
\begin{split}
\langle \mathcal{B}_f (p_f) | V_\mu - A_\mu | \mathcal{B}_i (p_i) \rangle &= \bar{u}_f^\nu
\left[ \left( \bar{f}_1(q^2) g_{\nu\mu} + \bar{f}_2(q^2) p_{i\nu} \gamma_\mu + \bar{f}_3(q^2) p_{i\nu}p_{f\mu} \right) \gamma_5 \right. \\
&\hspace{0.4in} \left.
- \left( \bar{g}_1(q^2) g_{\nu\mu} + \bar{g}_2(q^2) p_{i\nu} \gamma_\mu + \bar{g}_3(q^2)p_{i\nu}p_{f\mu} \right) \right]
u_i,
\end{split}
\label{eq:FFdef}
\end{equation}
with the same convention as in \cite{Cheng:1996cs}. In Appendix \ref{app:FFCon}, we summarize different conventions and provide correspondences among them.

\section{Form Factors in the Quark Model}
\label{sec:FF}

In this section, we comprehensively investigate non-perturbative $\frac{1}{2}^+ \to \frac{3}{2}^+$ baryon transition form factors within the non-relativistic constituent quark model to
 incorporate several types of decays consistently.
To describe a spin-$\frac{3}{2}$ particle in the final state, we introduce the Rarita-Schwinger vector spinor $u^{\mu}$ \cite{Lurie1969, Cheng:1996cs}. Its components for $\frac{3}{2}$, $\frac{1}{2}$, $-\frac{1}{2}$, and $-\frac{3}{2}$ spins are given by:
\begin{equation}
\begin{split}
&u^{\mu}_1=(0,\vec{\epsilon}_1u_\uparrow)\;,\\
&u^{\mu}_2=\left(\sqrt{\frac{2}{3}}\frac{|\vec{p}|}{m}u_\uparrow,\frac{1}{\sqrt{3}}\vec{\epsilon}_1u_\downarrow-\sqrt{\frac{2}{3}}\frac{E}{m}\vec{\epsilon}_3u_{\uparrow}\right)\;,\\
&u^{\mu}_3=\left(\sqrt{\frac{2}{3}}\frac{|\vec{p}|}{m}u_\downarrow,\frac{1}{\sqrt{3}}\vec{\epsilon}_2u_\uparrow-\sqrt{\frac{2}{3}}\frac{E}{m}\vec{\epsilon}_3u_{\downarrow}\right)\;,\\
&u^{\mu}_4=(0,\vec{\epsilon}_2u_\downarrow)\;,\\
\end{split}
\label{RS-spin}
\end{equation}
where $\vec{p}$ is the momentum of the final state baryon along its $z$-axis, $\vec{\epsilon}_i$ are polarized vectors, 
and $u$ is the spinor for a spin-$\frac{1}{2}$ particle, giving
\begin{equation}
\vec{\epsilon}_1=\frac{1}{\sqrt{2}}
\begin{pmatrix}
1\\
i\\
0\\
\end{pmatrix}\;,\;
\vec{\epsilon}_2=\frac{1}{\sqrt{2}}
\begin{pmatrix}
1\\
-i\\
0\\
\end{pmatrix}\;,\;
\vec{\epsilon}_3=\frac{1}{\sqrt{2}}
\begin{pmatrix}
0\\
0\\
1\\
\end{pmatrix};
\qquad
{u}(p)=
\begin{pmatrix}
1\\
\frac{\vec{\sigma}\cdot\vec{p}}{E+m}\\
\end{pmatrix} \chi,
\end{equation}
with the two-component Pauli spinor $\chi$ denoting spin up and down. In the rest frame of the parent baryon, the vanishing momentum of the initial particle implies the null lower component of $u_i$, while we have the relation $\vec{p}_f=-\vec{q}$ in the lower component of $u_f$.

 Now with explicit spinors, after
expanding the right-handed side of 6 typical independent equations, among all the 64 ones in the form factor definition Eq. (\ref{eq:FFdef}), we  obtain
\begin{align}
\begin{split}
\langle \mathcal{B}_f(1/2)|V_0|\mathcal{B}_i(1/2)\rangle&=\sqrt{\frac{2}{3}}\frac{|\vec{p}_f|^2}{m_f}\left[-\frac{1}{2m_f}\bar{f}_1+\frac{m_i}{2m_f}\bar{f}_2-\frac{m_i}{2}\bar{f}_3\right]\\
\langle \mathcal{B}_f(1/2)|A_0|\mathcal{B}_i(1/2)\rangle&=\sqrt{\frac{2}{3}}\frac{|\vec{p}_f|}{m_f}\left[\bar{g}_1+\bar{g}_2m_i+\bar{g}_3m_im_f\right]\\
\langle \mathcal{B}_f(3/2)|V_x|\mathcal{B}_i(1/2)\rangle&=\frac{|\vec{p}_f|}{\sqrt{2}}\frac{1}{2m_f}\bar{f}_1\\
\langle \mathcal{B}_f(3/2)|A_x|\mathcal{B}_i(1/2)\rangle&=-\frac{1}{\sqrt{2}}\bar{g}_1\\
\langle \mathcal{B}_f(1/2)|V_x|\mathcal{B}_i(-1/2)\rangle&=\frac{|\vec{p}_f|}{\sqrt{6}}\frac{1}{(2m_f)}\bar{f}_1+\sqrt{\frac{2}{3}}\frac{|\vec{p}_f|m_i}{m_f}\bar{f}_2\\
\langle \mathcal{B}_f(1/2)|A_x|\mathcal{B}_i(-1/2)\rangle&=-\bar{g}_1\frac{1}{\sqrt{6}}+\sqrt{\frac{2}{3}}\frac{|\vec{p}_f|^2m_i}{2m_f^2}\bar{g}_2\;
\end{split}\label{eq:FF-expan}
\end{align}
by choosing the $x$ component of the (axial-)vector part.

 Evidently, the six form factors can be expressed as combinations of the non-perturbative matrix elements on the left-hand side of Eq. (\ref{eq:FF-expan}). 
 We will investigate these non-perturbative parameters within the framework of the nonrelativistic constituent quark model to quantitatively understand the dynamics, given the absence of first-principle calculations. Utilizing the baryon wave function outlined in \cite{Zeng:2022egh}, a straightforward calculation, detailed in Appendix B, yields
\begin{align}
\begin{split}
&\langle \mathcal{B}_f(1/2)|V_0|\mathcal{B}_i(1/2)\rangle=N(I)|_{(\frac{1}{2},\frac{1}{2})}I_H\;,\qquad\qquad\;\;\;\langle \mathcal{B}_f(1/2)|A_0|\mathcal{B}_i(1/2)\rangle=\frac{|\vec{p}_f|}{2m_f}N(\sigma_z)|_{(\frac{1}{2},\frac{1}{2})}Z\;,\\
&\langle \mathcal{B}_f(3/2)|V_x|\mathcal{B}_i(1/2)\rangle=\frac{|\vec{p}_f|}{2m_f}N(\sigma_x)|_{(\frac{3}{2},\frac{1}{2})}X\;,\qquad\;\langle \mathcal{B}_f(3/2)|A_x|\mathcal{B}_i(1/2)\rangle=N(\sigma_x)|_{(\frac{3}{2},\frac{1}{2})}I_H\;,\\
&\langle \mathcal{B}_f(1/2)|V_x|\mathcal{B}_i(-1/2)\rangle=\frac{|\vec{p}_f|}{2m_f}N(\sigma_x)|_{(\frac{1}{2},-\frac{1}{2})}X\;,\\
&\langle \mathcal{B}_f(1/2)|A_x|\mathcal{B}_i(-1/2)\rangle=N(\sigma_x)|_{(\frac{1}{2},-\frac{1}{2})}(I_H-\frac{|\vec{p}_f|^2}{2m_f^2}Y)\;,
\end{split}
\label{eq:FF-matrix}
\end{align}
in which the spin-flavor factors $N(A)|_{(s_f,s_i)}(A=I,\sigma_x)$ are evaluated to be
\begin{equation}
\begin{split}
&N(I)|_{(\frac{1}{2},\frac{1}{2})}=0\;,\qquad\qquad N(\sigma_z)|_{(\frac{1}{2},\frac{1}{2})}=\frac{-2\sqrt{2}}{3}\;,\\
&N(\sigma_x)|_{(\frac{3}{2},\frac{1}{2})}=\sqrt{\frac{2}{3}}\;,\qquad N(\sigma_x)|_{(\frac{1}{2},-\frac{1}{2})}=\frac{\sqrt{2}}{3}\;,
\end{split}
\label{eq:spinflavornum}
\end{equation}
with particular third-component spin of initial baryon $s_i$ and final $s_f$,  
and the auxiliary functions
$X,Y,Z,Z',I_H$ given as
\begin{equation}
\begin{split}
&X=I_H\left[1+\frac{2m\alpha_{\lambda f}^2}{m_q(\alpha_{\lambda i}^2+\alpha_{\lambda f}^2)}+\frac{2m\alpha_{\lambda i}^2}{m_Q(\alpha_{\lambda i}^2+\alpha_{\lambda f}^2)}\right]\;,\\
&
Y=-I_H\left[\frac{2m^2\alpha_{\lambda i}^4}{m_qm_Q(\alpha_{\lambda i}^2+\alpha_{\lambda f}^2)^2}+\frac{m\alpha_{\lambda i}^2}{(\alpha_{\lambda i}^2+\alpha_{\lambda f}^2)m_Q}\right]\;,
\\
&Z=I_H\left[1+\frac{2m\alpha_{\lambda f}^2}{m_q(\alpha_{\lambda i}^2+\alpha_{\lambda f}^2)}-\frac{2m\alpha_{\lambda i}^2}{m_Q(\alpha_{\lambda i}^2+\alpha_{\lambda f}^2)}\right]\;,\\
&Z'=Z-Y-2I_H\;,\\
&I_H=\left(\frac{2\alpha_{\lambda i}\alpha_{\lambda f}}{(\alpha_{\lambda i}^2+\alpha_{\lambda f}^2)}\right)^{3/2}\exp\left[\frac{2m^2|\vec{q}|^2}{m_f(\alpha_{\lambda i}^2+\alpha_{\lambda f}^2)}\right],
\end{split}
\label{eq:XYZ}
\end{equation}
originate from integrals of baryon spatial wave functions, which are 
consistent with
the general form given in \cite{Pervin:2006ie}.
These explicit spatial integrals, which were absent from the earlier work \cite{Cheng:1996cs}, reveal the underlying dynamics of baryons.
The two types of parameters on which the spatial integrals rely are masses
(heavy (light) quark mass $m_Q$ ($m_q$) involving weak interaction, spectator quark mass $m$ and final baryon mass $m_f$) 
and harmonic oscillator parameter $\alpha_\lambda$ ($\alpha_\rho$ depends on $\alpha_\lambda$;  $\alpha_{\lambda i}$
($\alpha_{\lambda f}$) stands for initial (final) baryon parameter).
In this work, the naturally accommodated condition $\alpha_{\rho i}=\alpha_{\rho f}$,  due
to the identical $\rho$-mode shared by the initial particle $\Omega_c$ and final particle $\Omega, \Xi^{-}$,   further simplifies Eq. \eqref{eq:XYZ} and hence leads to a cancellation of
$\alpha_{\rho}$ dependence.\footnote{
The relation between $\lambda$-mode and $\rho$-mode 
presented as Eq. \eqref{eq:lamrho} is also taken into account. }

In the zero recoil limit ($q^2=q^2_{\rm{max}}$) , the FFs can be solved combining Eq. (\ref{eq:FF-expan}) and (\ref{eq:XYZ}), giving
\begin{equation}
\begin{split}
&\bar{f}_1=\frac{2}{\sqrt{3}}X\;,\qquad\bar{f}_2=\frac{1}{\sqrt{3}m_i}X\;,\qquad\bar{f}_3=\frac{-1}{\sqrt{3}m_im_f}X\\
&\bar{g}_1=-\frac{2}{\sqrt{3}}I_H\;,\qquad\bar{g}_2=-\frac{1}{\sqrt{3}m_i}Y\;,\qquad\bar{g}_3=-\frac{Z'}{\sqrt{3}m_im_f}\;,
\end{split}
\label{eq:FF}
\end{equation}
with $\mathcal{B}_i$ ($\mathcal{B}_f$) mass  $m_i$($m_f$) and the three-momentum $ |\vec{q}|=0$. 
Specifically, the uncertainties in all six form factors, and consequently in the factorizable amplitudes, 
arise solely from quark masses, owing to the cancellation involving $\alpha_\rho$ in the ratios 
shown in Eq. (\ref{eq:XYZ}).
In our practical calculation, we need the evolutions of form factors with  respect to $q^2$ to 
estimate physical observables. In this work, we follow the parametrization \cite{Fakirov1978}
\begin{equation}
f_i(q^2)=\frac{f_i(0)}{(1-q^2/m_V^2)^2}\;,\qquad g_i(q^2)=\frac{g_i(0)}{(1-q^2/m^2_A)^2}\;,
\label{eq:evo}
\end{equation}
adopted in our previous series of works 
\cite{Cheng:2018hwl,Zou:2019kzq,Cheng:2020wmk,Meng:2020euv,Cheng:2022kea, Hu:2020nkg,Zeng:2022egh}
to capture the typical evolution behaviors,
in which
the pole masses are taken as
$m_V=2.11~\text{GeV}, m_A=2.54~\text{GeV}$.


\section{Numerical results and discussions }
\label{sec:num}



\begin{table}[!t]
	\centering
	\caption{
	Input parameters adopted in this work. 
	Parameters listed in the upper entries without a reference are taken from PDG \cite{PDG:2024}, 
	while those in the lower entries 
	are from \cite{Wang:2017kfr}\footnote{
		The central values are quoted from \cite{Wang:2017kfr}, while uncertainties here
        are imposed for illustration of error analysis.
		 }.}
	\begin{tabular}{cccc}
		\hline\hline
		Parameters& Values& Parameters & Values \\
		\hline
		   $ m_{\Omega_c^0} $&$2.695$\,GeV&$ m_{\Omega^-} $& $1.672$\,GeV\\
            $ m_{\Xi^0} $&$1.314$\,GeV&$ m_{\Xi^-} $& $1.321$\,GeV\\
		$ a_1 $&$ 1.257 $  \cite{Cheng:2018hwl} &$ \tau_{\Omega_c^0} $&$ (268\pm26)\,\text{fs} $\cite{LHCb:2018nfa}\\
	        $ V_{cs} $&$ 0.9735 $&$ V_{ud} $&$ 0.9743 $\\
		$ V_{cd} $&$ -0.221 $&$ V_{us} $&$ 0.2251 $\\
		$f_\pi$ & $130.4(2)\,{\rm{MeV}}$ \cite{ExtendedTwistedMass:2021qui} 
		& $f_\rho$& $216\,{\rm{MeV}}$ \cite{Maris:1999nt}\\
		
            \hline
           
		$ m_u $& $(0.33\pm 0.05)$\,GeV  &$ m_d $& $(0.33\pm 0.05)$\,GeV  \\
		$ m_{s} $&$(0.45\pm 0.05)$\,GeV&$ m_c $& $(1.48\pm 0.10)$\,GeV\\
		$ \alpha_{\rho} $& $(0.325\pm 0.025)$\,GeV 
		\footnote{
		Due to the difference in definition conventions,
		the value adopted here is smaller by a factor of $\sqrt{2}$ compared with the one defined in \cite{Wang:2017kfr},  where $\alpha_\rho = 0.44\, \text{GeV}.$ The slight deviation in central value can be accounted
		by the imposed error.
		 }
		& \\
		\hline\hline
	\end{tabular}
	\label{tab:inputs}
\end{table}

Before delving into detailed numerical calculations, interpreting the results, and discussing input sensitivities, we first summarize the input parameters utilized in the following analyses, presented in Table \ref{tab:inputs}. Broadly categorized, these input parameters fall into two groups based on their sensitivity to the quark model. The first group comprises parameters that are largely physical, already determined either by experimental measurements or theoretical calculations. These include the masses and lifetimes of baryons, CKM matrix elements, and effective Wilson coefficients. Notably, the effective Wilson coefficient employed in this study is set to $a_1=1.257$, determined by fixing $N_{\rm{eff}}\approx 7$ as extracted from $\Lambda_c \to p \phi$ \cite{BESIII:2016ozn,Cheng:2018hwl}. The second group consists of parameters such as quark masses and the harmonic oscillator parameter $\alpha_\rho$, which are model-dependent. In the context of the constituent quark model, the central values of light quark masses are chosen as $m_u=m_d=m_l=0.33\,{\rm{GeV}}$, while the masses of strange and charm quarks are assigned as $0.45\,{\rm{GeV}}$ and $1.48\,{\rm{GeV}}$, respectively \cite{Wang:2017kfr}.



For each baryon, it appears that two harmonic oscillator parameters, $\alpha_\rho$ and $\alpha_\lambda$, are required to describe its corresponding excited mode. However, as illustrated in \cite{Zeng:2022egh}, these two strength parameters are interconnected via
\begin{equation}
\alpha_\lambda = \left[ \frac{4m_3(m_1+m_2)^2}{3m_1 m_2(m_1+m_2+m_3)}\right]^{\frac14} \alpha_\rho,
\label{eq:lamrho}
\end{equation}
where the paired quarks $q_{1,2}$ correspond to their masses $m_{1,2}$, resulting in only one independent oscillator strength for each baryon. 
Since the baryons $\Omega_c$, $\Omega$, and $\Xi^{0,-}$ share the same $\rho$-mode, consisting of a pair of strange quarks, it is reasonable to use the relation $\alpha_\rho(\Omega_c)=\alpha_\rho(\Omega)=\alpha_\rho(\Xi^{0,-})$. 
The fact that $\alpha_{\rho i}=\alpha_{\rho f}$ and 
$\alpha_{\lambda i}=[(2m_s+m_c)/(3m_c)]^{1/4}\alpha_{\lambda f}$
 simplifies our analysis by reducing the number of free parameters to one, $\alpha_\rho$, for all processes discussed in this work. In the subsequent numerical evaluations, we adopt the value $(0.325\pm 0.025)\,\rm{GeV}$ for $\alpha_\rho$ \cite{Wang:2017kfr}, 
 leading to $\alpha_{\lambda}(\Omega_c^0)=(0.438 \pm 0.034)\,\rm{GeV}$, $\alpha_{\lambda}(\Omega^-) =(0.375 \pm 0.029)\,\rm{GeV}$, and $\alpha_{\lambda}(\Xi^0, \Xi^-)=(0.355\pm 0.027)\,\rm{GeV}$.
All the inputs relevant to the numerical analyses in this work are presented in Table \ref{tab:inputs}.

\subsection{Form factors}
\label{subsec:FF}


\begin{figure}[t!]
	\centering  
	\includegraphics[width=1.0\linewidth]{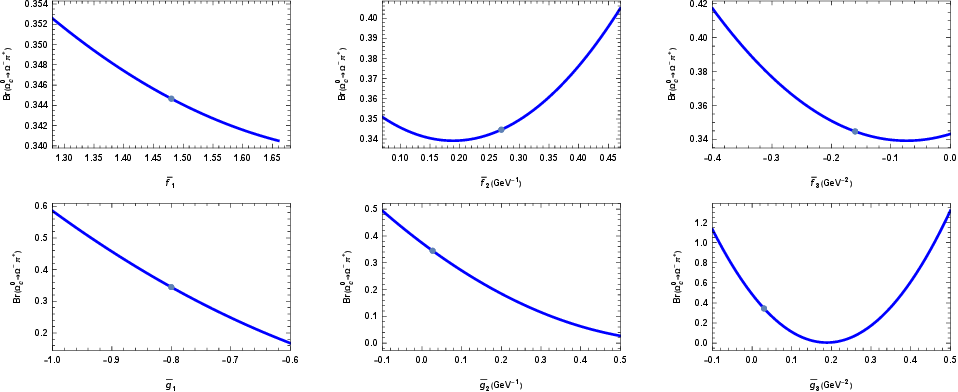}
    \caption{The form factor dependence of  branching fraction for the decay $\Omega_c\to \Omega^- \pi^+$. }  
    \label{fig:FF-BR}
\end{figure}

The decay mode $\Omega_c\to\Omega^-\pi^+$ holds significant importance in experiments as it serves as a benchmark mode. Moreover, this channel holds a special place in theory as it receives pure factorizable contributions, relying on the $\Omega_c\to \Omega^-$ transition form factors. Before delving into the detailed dynamics of these form factors, a preliminary examination of their impact on the branching fraction proves beneficial. It is understood that in $\frac12^+\to \frac12^+$ decays, the branching fractions are predominantly governed by $f_1$ and $g_1$, while the contributions from the others can be overlooked. However, in the case of $\frac12^+\to \frac32^+$ transitions, each form factor contributes significantly. In each plot of Fig. \ref{fig:FF-BR}, marked by the point at the physical scale, although the individual behaviors vary, the contributions to the branching fraction from all 6 form factors are of a similar magnitude.

\begin{figure}[t!]
	\centering  
	\includegraphics[width=1.0\linewidth]{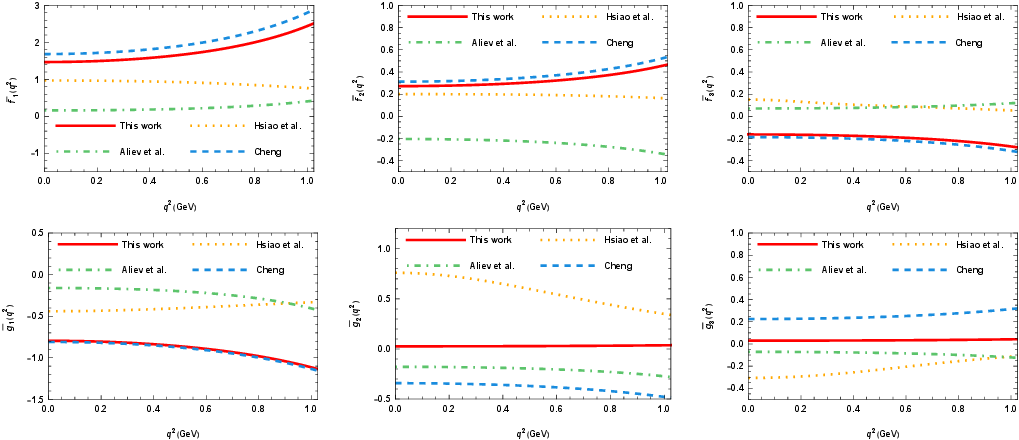}
    \caption{The evolution of form factors compare with several methodoligies, including the early NRQM \cite{Cheng:1996cs},  
    light cone sum rules \cite{Aliev:2022gxi} and light-front quark model \cite{Hsiao:2020gtc}.  }
    \label{fig:FF-behavior}
\end{figure}

In the context of the non-relativistic constituent quark model, we illustrate the individual behaviors of the 6 form factors of the $\Omega_c^0\to \Omega^-$ process in 
Fig. \ref{fig:FF-behavior}, which primarily 
depend on the constituent quark mass and the harmonic oscillator strength $\alpha_\rho$. Among the 6 form factors, $\overline{f}_1$, $\overline{f}_2$, $\overline{g}_2$, and $\overline{g}_3$ exhibit positive values, while $\overline{f}_3$ and $\overline{g}_1$ are negative. The predicted behaviors presented by other groups, with convention transferred, are also included for comparison. Although the real integral of the spatial wave function was not considered in the previous non-relativistic quark model calculation \cite{Cheng:1996cs}, our results demonstrate consistent and similar behaviors for $\overline{f}_1$, $\overline{f}_2$, $\overline{f}_3$, and $\overline{g}_1$. 
The discrepancies are more apparent when comparing our results with calculations from the light-cone sum rule \cite{Aliev:2022gxi} and the light-front quark model \cite{Hsiao:2020gtc}. For instance, regarding $\overline{f}_1$, our results agree with the light-cone sum rule in terms of evolution behavior but differ in size. In contrast, when comparing with the light-front quark model prediction, the size difference is up to two times, while the evolution behaviors are opposite. Hence, further scrutiny of these form factors from experimental data or lattice QCD simulations is necessary and anticipated.

\subsection{Branching fractions }
\label{subsec:BR}

Without a clear understanding of the reference mode $\Omega_c\to \Omega^- \pi^+$, the predictions of other modes lack conviction. Therefore, our top priority here is to provide an explicit numerical prediction for this benchmark channel.

\begin{table}[b!]
\center
\caption{Predictions of absolute and relative branching fractions, together with
comparisons among different theoretical groups.
Only errors for $\Omega_c\to \Xi^- \pi^+$ has been denoted, originated
from its component of non-factorizable contribution. } 
\label{tab:Brs}
\vspace{-6pt}
\setlength{\tabcolsep}{1mm}
\resizebox{\textwidth}{!}{\begin{threeparttable}
\begin{tabular}{cccccccccc}
\hline\hline
\multicolumn{2}{c}{}                   
&  This  work & Cheng
 \cite{Cheng:1996cs}& Wang et al.\cite{Wang:2022zja} &  Gutsche et al.\cite{Gutsche:2018utw}&  Hsiao et al.\cite{Hsiao:2020gtc}  &Aliev et al.\cite{Aliev:2022gxi} &Liu \cite{Liu:2023dvg}\\ \hline
\multicolumn{2}{l}{$\Omega_{c}^{0}\rightarrow\Omega^{-}\pi^+$}                 
  &  $(3.43\pm 0.48)$\%    &  4.19\%  &  1.1\%  & 0.2\%  & 0.51\%& 2.9\% & 1.88\%   \\
\multicolumn{2}{l}{$\Omega_{c}^{0}\rightarrow\Omega^{-}\rho^+$}                   
&   $(18.30\pm 1.94)$\% & 15.08\%   &    & 1.9\% & 1.44$\pm$0.04\% &6.3\%&\\
\multicolumn{2}{l}{$\Omega_{c}^{0}\rightarrow\Xi^{-}\pi^+$}                   
&   $(0.54{\pm 0.18})\%$ &  &    & & &&0.94\% \\
\multicolumn{2}{l}{$\Omega_{c}^{0}\rightarrow\Omega^{-}e^+\nu_e$}                   
&   $(4.06\pm 0.48)$\% &  &    & & 0.54$\pm$0.02\% &2.06\%&2.54\%\\
\multicolumn{2}{l}{$\Omega_{c}^{0}\rightarrow\Omega^{-}\mu^+\nu_\mu$}                   
&   $(3.81\pm 0.44)$\% &  &    & & 0.50$\pm$0.02\% &1.96\%\\
 \hline
\multicolumn{2}{l}{$R({\Omega^{-}\rho^+})$}                        
&  $5.33\pm 0.94$  &   3.60  &   & 9.5 &  2.8$\pm$0.4  &2.18&\\
\multicolumn{2}{l}{$R({\Xi^{-}\pi^+})$}                        
&  $0.16\pm 0.06$  &     &   &  &   &  &0.5&\\
\multicolumn{2}{l}{$R({\Omega^{-}e^+\nu_e})$}                        
&  $1.18\pm 0.22$&     &   &  &  1.1$\pm$0.2 & 0.71&1.35\\
\multicolumn{2}{l}{$R({\Omega^{-}\mu^+\nu_\mu})$}                        
&  $1.11\pm 0.20$  &     &   &  &  1.059 & 0.68&\\
\multicolumn{2}{l}{$R({\Omega^{-}e^+\nu_e})/R({\Omega^{-}\mu^+\nu_\mu})$}                     &  $1.07\pm 0.18$  &     &   &  &  1.08 &1.04 &\\
\hline\hline
\end{tabular}\footnotesize
\end{threeparttable}}
\end{table}

\subsubsection{The benchmark channel $\Omega_c\to \Omega^- \pi^+$ 
and $\Omega_c\to \Omega^- \rho^+$ }

The decay mode $\Omega_c\to \Omega^- \pi^+$ holds special significance as it primarily involves purely factorizable contributions described by baryon transition form factors. By setting up the kinematics for $\frac12^+\to \frac32^+ + 0^-$ and utilizing the analytically calculated form factors $\bar{f}_i(m_P^2)$ and $\bar{g}_i(m_P^2)$, we can compute its branching fraction. Our predicted value of 
\begin{equation}
\mathcal{B}(\Omega_c^0\to \Omega^- \pi^+)=(3.43\pm0.48)\%
\end{equation} 
closely aligns with the early nonrelativistic quark model (NRQM) prediction of $4.19\%$, but notably exceeds the predictions of other models. Specifically, constituent quark model (CQM) results \cite{Wang:2022zja} hover around the percentage range, whereas predictions are even smaller in both the covariant confined quark model (CCQM) \cite{Gutsche:2018utw} and the light-front quark model (LFQM) \cite{Hsiao:2020gtc}.

The decay $\Omega_c\to \Omega^- \rho^+$ shares similarities with the benchmark mode, as its amplitude predominantly consists of purely factorizable contributions. However, the kinematics are more complicated due to the presence of the final vector meson. Despite this complexity, we calculate a branching fraction of $18.39\%$, 
which closely resembles the early NRQM prediction of $15.08\%$, but is approximately an order of magnitude larger than the predictions of CCQM \cite{Gutsche:2018utw} and LFQM \cite{Hsiao:2020gtc}.

To date, directly measuring the benchmark mode remains challenging. However, some ratios between other modes and the reference mode are available. Notably, only Belle has reported a lower bound of $R(\Omega^- \rho^+)>1.3$, which aligns with the predictions of all theoretical models.

\subsubsection{Non-leptonic decay $\Omega_c\to \Xi^- \pi^+$ }
\begin{table}[!t]
\center
\caption{The predicted $S$- and $P$-wave amplitude (in units of $10^{-2}G_F^2$ GeV) of $\Omega_{c}^{0}\rightarrow\Xi^{-}\pi^+$ decay.} 
\label{tab:Xipi}
\vspace{-3pt}
\setlength{\tabcolsep}{3mm}
\begin{threeparttable}
\resizebox{\textwidth}{!} 
{
\begin{tabular}{cccccccccc}
\hline\hline
\multicolumn{2}{c}{}                   
&  $A^{\text{fac}}$ & $A^{\text{nf}}$& $A^{\text{tot}}$ &   $B^{\text{fac}}$ & $B^{\text{nf}}$& $B^{\text{tot}}$ &$\mathcal{B}$\\ \hline
\multicolumn{2}{l}{$\Omega_{c}^{0}\rightarrow\Xi^{-}\pi^+$}                 
  &  $-0.950\pm0.094$   &  $-2.233 \pm 0.549 $  &  $-3.183 \pm 0.557 $  & $-1.493\pm0.150 $  
  & $1.481 \pm 0.343$& $-0.012 \pm0.373 $ &  $(0.535  \pm0.179) \%$   \\
\hline\hline
\end{tabular}
}
\end{threeparttable}
\end{table}

The mode  $\Omega_c\to \Xi^- \pi^+$ is a typical non-leptonic decay since it receives both factorizable and non-factorizable contributions.
In the framework of topological diagram approach by combining the pole model at hadron level and MIT bag model at quark level, 
$\Omega_c$ decays have been studied
systematically in  \cite{Hu:2020nkg}.  To reduce the uncertainty of  quark model estimation of non-perturbative parameters, it is
valuable to investigate them in a separated approach. General speaking,
the momentum
carried by final state particle in the decays of charm system is with smaller velocity comparing with those decayed from B system, thus a choice of
 non-relativistic quark model is reasonable. 
 By collecting  necessary formulae for factorizable and non-factorizable amplitudes directly from  \cite{Hu:2020nkg}
 \begin{equation}
\begin{split}
 &A^{\text{fac}}=
 \frac{G_F}{\sqrt{2}} a_1 V^\ast_{ud} V_{cd} f_\pi (m_{\Omega_c}-m_{\Xi^-}) f_1,
 \\
 &B^{\text{fac}}= 
 - \frac{G_F}{\sqrt{2}} a_1 V^\ast_{ud} V_{cd} f_\pi (m_{\Omega_c}+m_{\Xi^-}) g_1,
 \\
 &A^{\text{nf}}=\frac{1}{f_\pi} a_{\Xi^0 \Omega_c^0}, \\
 &B^{\text{nf}}=\frac{1}{f_\pi}\left( g^{A(\pi^+)}_{\Xi^- \Xi^0} 
 \frac{m_{\Xi^-}+m_{\Xi^0}}{m_{\Omega_c^0}-m_{\Xi^0}}
 a_{\Xi^0 \Omega_c^0}\right), \\
\end{split}
\end{equation} 
 and numerically calculating corresponding non-perturbative parameters  
\begin{equation}
f_1 =0.276\pm0.028 ,  \quad  g_1 =-0.148\pm0.015, \quad
g_{\Xi^-\Xi^0}^{A(\pi^+)}= -0.334\pm0.044, 
\quad
a_{\Xi^0\Omega_c^0}=(-3.148\pm 0.715)\times10^{-3}
\end{equation}
we  predict its
 branching fraction as
\begin{equation}
\mathcal{B}(\Omega_c \to \Xi^-\pi^+) = (5.35\pm 1.79)\times10^{-3}\;.
\end{equation}
For the $\alpha_\rho$ cancellation in FFs,
the dominated error originated from oscillator parameter  only affects
non-factorizable contribution. Therefore only the errors of  four-quark operator matrix element
and hence branching fraction
have been calculated.
Compared with our previous MIT bag model calculation, in which $f_1 =0.25 ,   g_1 =-0.12,
g_{\Xi^-\Xi^0}^{A(\pi^+)}= -0.217, 
a_{\Xi^0\Omega_c^0}=-4.54\times10^{-4}, $ and hence 
$\mathcal{B}(\Omega_c \to \Xi^-\pi^+) = 9.34\times10^{-3}$  \cite{Hu:2020nkg}, the decrease of branching fraction is mainly due to an enhancement of matrix element of four-quark operator $a_{\Xi^0 \Omega_c^0}$.
Similar enhancement of matrix element, and hence enhancement of non-factorizable contribution, was also observed in the study of
decay of doubly charmed baryon decays
\cite{Zeng:2022egh}, by comparing MIT bag model and NR constitute quark model.
We display the contributed components of $\Omega_c \to \Xi^-\pi^+$  in 
Table \ref{tab:Xipi} for more details.
Although it is still challenging to give a direct measurement of such a decay, 
the calculated relative ratio 
\begin{equation}
R(\Xi^- \pi^+) =0.156\,\pm\, 0.057
\end{equation}
is consistent well with the recent measured value
$0.158\pm0.011$
 reported by LHCb \cite{LHCb:2023fvd}.


\subsubsection{Semi-leptonic decays  $\Omega_c\to \Omega^- \ell^+ \nu_\ell$}

For the semi-leptonic decay, the leptons in the final state is specific to electron and muon.
Obviously in the massless limit of leptons, the branching fraction is identical
since it only receives contribution from $H_V$ according to Eq. (\ref{eq:sl}).
In practice, the lepton mass effect accounts, leading to close but
slightly deviated 
branching fractions for $\Omega_c\to \Omega^- e^+ \nu_e$ and
$\Omega_c\to \Omega^- \mu^+ \nu_\mu$.
With the prepared FFs $\overline{f}_i$ and $\overline{g}_i$, the
branching fractions of
 semi-leptonic decays 
 are calculated to be
 \begin{equation}
 \mathcal{B}(\Omega_c \to \Omega^- e^+ \nu_e) = (4.06\pm0.48)\%,\quad
 \mathcal{B}(\Omega_c \to \Omega^- \mu^+ \nu_\mu) = (3.81\pm0.44)\%,
 \end{equation}
which are around one order of magnitude larger than the predictions
in  \cite{Hsiao:2020gtc} and twice larger than calculation in light-cone sum rule \cite{Aliev:2022gxi}.
Although so far no direct discrimination from experiment is available, 
the relative ratios to reference mode can still provide supportive information.
Results in our calculation,
\begin{equation}
R(\Omega^- e^+ \nu_e)= 1.18\pm0.22,\quad
R(\Omega^- \mu^+ \nu_\mu)= 1.11\pm0.20,
\end{equation}
are slight smaller than \cite{Liu:2023dvg}, close to the one obtained in  \cite{Hsiao:2020gtc}, but quite larger than \cite{Aliev:2022gxi} among the theoretical groups.
Our predictions,
On the other hand, though are smaller than 
Belle measurement $R(\Omega^- e^+ \nu_e)= 1.98\pm 0.15,
R(\Omega^- \mu^+ \nu_\mu)= 1.94+0.21$, are consistent with
recent ALICE reported value 
$R(\Omega^- e^+ \nu_e)= 1.12\pm 0.35$.
A further discrimination by more precise measurement in experiment
or calculation from first principle in theory is thus highly anticipated.


\section{Concluding remarks}
\label{sec:con}

Inspired by the recent LHCb and ALICE measurement of $\Omega_c$ decays,
we study  its semi-leptonic and non-leptonic decay processes in this work.
Our main results are as follows. 

\begin{itemize}

\item 
Branching fractions of $\Omega_c\to \Omega P$,
$\Omega_c\to \Omega V$ and $\Omega_c\to \Omega \ell^+ \nu_\ell$ 
all receive contributions from 
$\frac12 ^+\to \frac32 ^+$ transition form factors. 
Conventions of  these FFs are unified. 
Unlike $\frac12 ^+\to \frac12 ^+$ decays, in which $f_1$ and $g_1$ give dominated contributions to branching fraction,
all the 6 FFs provide comparable components to the branching fraction as displayed in Fig. \ref{fig:FF-BR}.
 
\item In the framework of NRQM, we performed an analytical calculation
of 6 FFs  related to $\Omega_c^0 \to \Omega^- $. It is interesting to note that all the FFs depends
only on constitute quark masses and one oscillation parameter $\alpha_\rho$.
We make a comparison with other groups for the estimation of FFs numerically. As shown in Fig. \ref{fig:FF-behavior},
deviations exist among different methods and further scrutiny is required in the future. 

\item We estimated absolute and relative branching fractions of types of decays, including
benchmark mode
$\Omega_c^0 \to \Omega^- \pi^+$, $\Omega_c^0 \to \Omega^- \rho^+$, $\Omega_c^0 \to \Xi^- \pi^+$, 
$\Omega_c^0 \to \Omega^- e^+ \nu_e$ and $\Omega_c^0 \to \Omega^- \mu^+ \nu_\mu$.  The results are presented
in Table \ref{tab:Brs}.

\item Although the above absolute branching fractions can not be measured directly, some of their relative ratios 
to benchmark mode have been measured recently.  
Within NRQM and taking quark mass and oscillator parameter $\alpha_\rho$ as inputs, our predictions
generally satisfy all the current avaliable experimental data.
Especially, 
our prediction of $R(\Xi^- \pi^+)$ 
agrees well with the recent 
LHCb 2023 measurement for the first time, which is evident from a combination of Table \ref{tab:exp} and \ref{tab:Brs}.
For the semi-leptonic decays, although the prediction of $R(\Omega^- e^+ \nu_e)$ and
$R(\Omega^- \nu^+ \nu_\mu)$ in this work has more than $5\sigma$ deviation from Belle 2022 result,
it is consistent well with recent result reported by ALICE 2024.

\end{itemize}

 The study of $\Omega_c$ offers valuable insights into charmed baryon physics and 
 opens up new opportunities to explore the dynamics at the charm scale. 
Further developments on $\Omega_c$  both in experiment and theory, including 
 first principle calculation from Lattice QCD, are highly anticipated.

 \vskip 1cm
 \noindent {\textbf{Note Added}}
 
\noindent All the authors contribute equally and they are co-first authors, while F. Xu is
the corresponding author.

\vskip 2.0cm \acknowledgments

We would like to thank Hai-Yang Cheng, Xiao-Rui Lyu, Jin-Lin Fu and Feng-Zhi Chen for valuable discussions. 
This research is supported by  the National Natural Science Foundation of China under Grant Nos. 12475095 and U1932104.


\appendix

\section{Conventions of form factors}
\label{app:FFCon}

For the baryon decay $\frac12^+ \to \frac32^+$,  generically there are $8$  form factors to describe the baryon transition under the $V-A$  
current.  One parameterization, see \cite{Gutsche:2018utw, Gutsche:2019iac},  is taken as 
\begin{equation}
\begin{split}
\langle \mathcal{B}_f(p_f)|V_{\mu}-A_{\mu}|\mathcal{B}_i(p_i)\rangle
&=\bar{u}_f^{\nu}\left[\left(F_1(q^2)g_{\nu\mu}+F_2(q^2)\frac{p_{i\nu}}{m_i}\gamma_{\mu}+F_3(q^2)\frac{p_{i\nu}p_{f\mu}}{m_i^2}+F_4(q^2)\frac{p_{i\nu}q_{\mu}}{m_i^2}\right)\gamma_5\right.\\
&\left.\;\;\;\;-\left(G_1(q^2)g_{\nu\mu}+G_2(q^2)\frac{p_{i\nu}}{m_i}\gamma_{\mu}+G_3(q^2)\frac{p_{i\nu}p_{f\mu}}{m_i^2}+G_4(q^2)\frac{p_{i\nu}q_{\mu}}{m_i^2}\right)\right]u_i.
\end{split}
\label{eq:FFcv2}
\end{equation}
The (axial-)vector current form factor $F_j$ ($G_j$) ($j=1,2,3,4$) depends on momentum transfer $q^2$, and $q$ is defined as $q=\revs{p_i-p_f} $.  
In our work, we take a convention with $6$ independent form factors (see Eq. (\ref{eq:FFdef})) which was adopted in \cite{Cheng:1996cs}.
In fact, by making use of the following relations, 
\begin{equation}
\begin{split}
&\bar{f}_1=F_1\;,\qquad\bar{f}_2=\frac{F_2}{m_i}\;,\qquad  p_{f\mu}\bar{f}_3=\frac{p_{f\mu} F_3+q_\mu F_4 }{m_i^2}\\
&\bar{g}_1=G_1\;,\qquad\bar{g}_2=\frac{G_2}{m_i}\;,\qquad p_{f\mu}\bar{g}_3=\frac{p_{f\mu} G_3+q_\mu G_4}{m_i^2}\;.\\
\end{split}
\label{eq:FFcorr1}
\end{equation}
one may easily understand the two parameterizations are equivalent. 

Here we display the difference and connection between parameterizations of FFs in terms of
Eq. (\ref{eq:FFcv2}) and Eq. (\ref{eq:FFdef}). Whether to incorporate $F_4$ and $G_4$ is the main crux.
Since they do not enter the contributed amplitudes of modes decaying into vector meson
in view of helicity amplitudes \cite{Gutsche:2018utw}, 
we only need to consider decays of $\frac12^+
\to \frac32^+ + 0^-$.  
Taking the $D$ term of Eq. (\ref{eq:pwa}) as an example, with
the help of relation $q\cdot \revs{p_f}=\frac12(m_i^2-m_f^2-q^2)=(m_iE_f- m_f^2)$ by incorporating 
on-shell condition $q^2=m_P^2$, we have
\begin{eqnarray}
D&= &\lambda a_1 f_P [\bar{f}_1 - (m_i+ m_f)\bar{f}_2 + (m_i E_f -m_f^2)\bar{f}_3]\nonumber\\
&=& \lambda a_1 f_P\left[ F_1 -\frac{m_i + m_f}{m_i} F_2 + \frac{p_f\cdot q F_3 +q^2 F_4}{m_i^2}\right]\nonumber\\
&=& \lambda a_1 f_P\left[ F_1 - F_2 \frac{m_i + m_f}{m_i} + F_3 \frac{m_i^2 - m_f^2 -q^2}{2m_i^2}
+ F_4 \frac{q^2}{m_i^2} \right],
\end{eqnarray}
where the corresponding relation Eq. (\ref{eq:FFcorr1}) has been used in the second equation.
Apparently,  as contributed components to  decay width,
our Eq. (\ref{eq:pwa}) is equivalent  to  corresponding terms in Eq. (43) of  \cite{Gutsche:2018utw}.

There is another set of convention 
in terms of 8 independent form factors adopted in \cite{Zhao:2018mrg,  Hsiao:2020gtc, Aliev:2022gxi, Lu:2023rmq, Pervin:2006ie},  giving 
\begin{equation}
\begin{split}
\langle \mathcal{B}_f^*(p_f)|V_{\mu}-A_{\mu}|\mathcal{B}_i(p_i)\rangle&= \bar{u}_f^{\nu}
\left[\left(f_1(q^2)\frac{p_{i\nu}}{m_i}\gamma_{\mu}+f_2(q^2)\frac{p_{i\nu}p_{i\mu}}{m_i^2}+f_3(q^2)\frac{p_{i\nu}p_{f\mu}}{m_im_f}+f_4(q^2)g_{\nu\mu}
\right)\gamma_5\right.\\
&\left.\;\;\;\;-\left(g_1(q^2)\frac{p_{i\nu}}{m_i}\gamma_{\mu}+g_2(q^2)\frac{p_{i\nu}p_{i\mu}}{m_i^2}+g_3(q^2)\frac{p_{i\nu}p_{f\mu}}{m_im_f}+g_4(q^2)g_{\nu\mu}\right)\right]u_i.
\end{split}
\end{equation}
By applying the following correspondence
\begin{equation}
\begin{split}
&\bar{f}_1=f_4\;,\quad \bar{f}_2=\frac{f_1}{m_i}\;,\quad \bar{f}_3=\frac{(f_2+f_3)}{m_im_f}\;, \\
&\bar{g}_1=g_4\;,\quad \bar{g}_2=\frac{g_1}{m_i}\;,\quad \bar{g}_3=\frac{(g_2+g_3)}{m_im_f}\;,   
\end{split}
\label{eq:FFcorr2}
\end{equation}
an equivalence to our Eq. (\ref{eq:FFdef}) can be established.

\section{Model calculations of form factors}
\label{app:FFcalc}

Taking $\bar{f}_1$ as an example, we show necessary details on the calculation of FFs in this section.
We shall start from the relations between matrix elements and form factors Eq. (\ref{eq:FF-matrix}) at the rest frame of the parent baryon. 
The quark operator $V_{\mu}=\bar{q}\gamma_{\mu}Q$ (or $A_{\mu}=\bar{q}\gamma_{\mu}\gamma_5Q$) will be 
expanded by
two-component 
Pauli spinor 
with
\begin{equation}
\begin{split}
Q=\begin{pmatrix}
1\\
\frac{\boldsymbol{\sigma}\cdot\textbf{p}_Q}{2m_Q}\\
\end{pmatrix}\chi_{\pm}\;,\qquad \bar{q}=\chi^\dagger_{\pm}\begin{pmatrix}
1,&\frac{\boldsymbol{\sigma}\cdot\textbf{p}_q}{2m_q}\\
\end{pmatrix}\;.
\end{split}
\end{equation}
To match specific components, we denote $\mathbf{p}_Q$ and $\mathbf{p}_q$ as the momenta of the heavy quark $Q$ and the light quark $q$, which are involved in the decay $Q \to q$ at the quark level. These momenta as parts of 
momenta of the initial and final baryons, satisfying $\mathbf{p}_i-\mathbf{p}_f=\mathbf{p}_Q-\mathbf{p}_q$.
In order to extract $\bar{f}_1$, we take the spatial component $x$  and
spin component ($s_i$,$s_f$)=($\frac{1}{2}$,$\frac{3}{2}$) in Eq. (\ref{eq:FF-matrix}), giving 
\begin{equation}
\begin{split}
\langle \mathcal{B}_f(3/2)|V_{x}|\mathcal{B}_i(1/2)\rangle&=\int d\textit{\textbf{p}}_{\rho i}d\textit{\textbf{p}}_{\lambda i}d\textit{\textbf{p}}_{\rho f}d\textit{\textbf{p}}_{\lambda f}\Psi^{*3/2}_{(\textit{\textbf{p}}_{\rho f},\textit{\textbf{p}}_{\lambda f})}\Psi^{1/2}_{(\textit{\textbf{p}}_{\rho i},\textit{\textbf{p}}_{\lambda i})}\langle q'_{3}q'_{2}q|\bar{q}\gamma_1Q|Qq_{2}q_{3}\rangle\\
&=\int d\textit{\textbf{p}}_{\rho i}d\textit{\textbf{p}}_{\lambda i}d\textit{\textbf{p}}_{\rho f}d\textit{\textbf{p}}_{\lambda f}\Psi^{*3/2}_{(\textit{\textbf{p}}_{\rho f},\textit{\textbf{p}}_{\lambda f})}\Psi^{1/2}_{(\textit{\textbf{p}}_{\rho i},\textit{\textbf{p}}_{\lambda i})}\langle q'_{3}q'_{2}|q_{2}q_{3}\rangle\langle q|\bar{q}\gamma_1Q|Q\rangle\\
&=N(\sigma_x)|_{(\frac{3}{2},\frac{1}{2})}\int d\textit{\textbf{p}}_{\rho i}d\textit{\textbf{p}}_{\lambda i}d\textit{\textbf{p}}_{\rho f}d\textit{\textbf{p}}_{\lambda f}\Psi^{*3/2}_{(\textit{\textbf{p}}_{\rho f},\textit{\textbf{p}}_{\lambda f})}(\frac{\textit{\textbf{p}}_{q}}{2m_q}-\frac{\textit{\textbf{p}}_{Q}}{2m_Q})\Psi^{1/2}_{(\textit{\textbf{p}}_{\rho i},\textit{\textbf{p}}_{\lambda i})}\\
&\;\;\;\;\times\delta^3(\textit{\textbf{p}}_{\rho i}-\textit{\textbf{p}}_{\rho f})\delta^3(\textit{\textbf{p}}_{\lambda i}-\textit{\textbf{p}}_{\lambda f}-\frac{2m_l}{m_i}\textit{\textbf{p}}_{f})\\
&=\frac{|\vec{p}_f|}{2\sqrt{2}m_f}\bar{f}_1\;.
\end{split}\label{eq:matrixFFC}
\end{equation}
In Eq. (\ref{eq:matrixFFC}), the spin-flavor factor is denoted as $N(\sigma_x)|_{(\frac{3}{2},\frac{1}{2})}$ and
 delta functions originate from Jacobi momentum (see \cite{Zeng:2022egh}).
 The baryon wave function $\Psi^{3/2}_{(\textit{\textbf{p}}_{\rho f},\textit{\textbf{p}}_{\lambda f})}$ 
 and $\Psi^{1/2}_{(\textit{\textbf{p}}_{\rho i},\textit{\textbf{p}}_{\lambda i})}$ 
 can be generally expressed in terms of Jacobi momentum $\textit{\textbf{p}}_{\rho}$ and $\textit{\textbf{p}}_{\lambda}$
\begin{align}
\Psi_{N,L,M_L}(\textit{\textbf{P}},\textit{\textbf{p}}_{\rho},\textit{\textbf{p}}_{\lambda})=\delta^3(\textit{\textbf{P}}-\textit{\textbf{P}}_c)\sum_{m}\langle LM_L|l_{\rho }m,l_{\lambda}M_L-m\rangle\psi_{n_{\rho}l_{\rho}m}(\textit{\textbf{p}}_{\rho})\psi_{n_{\lambda}l_{\lambda}(M_L-m)}(\textit{\textbf{p}}_{\lambda})\;,
\label{eq:BaryonWF}
\end{align}
associated with the quark wave function in momentum space
\begin{align}
\psi_{nLm}(\textit{\textbf{p}})=(i)^l(-1)^n\big{[}\frac{2n!}{(n+L+\frac{1}{2})!}\big{]}^\frac{1}{2}\frac{1}{\alpha^{L+\frac{3}{2}}}\exp\left[-\frac{\textit{\textbf{p}}^{2}}{2\alpha^{2}}\right]L_{n}^{L+\frac{1}{2}}(\frac{\textit{\textbf{p}}^{2}}{\alpha^2})\mathcal{Y}_{Lm}(\textit{\textbf{p}})\;.
\label{wf}
\end{align}
For a baryon state with the quantum number $J^P=\frac{3}{2}^+$ in this work, its orbital angular momentum is taken as $L=0$ to set the parity, yielding the wave function,
\begin{equation}
\begin{split}
\Psi^{3/2}_{(\mathbf{p}_{\rho f}, \mathbf{p}_{\lambda f})} = \psi_{000}(\mathbf{p}_{\rho f})\psi_{000}(\mathbf{p}_{\lambda f}) = \frac{1}{\sqrt{\pi^3}}\frac{1}{(\alpha_{\rho f}\alpha_{\lambda f})^{\frac{3}{2}}}\exp\left[-\frac{1}{2}\left(\frac{\mathbf{p}_{\rho f}^2}{\alpha_{\rho f}^2}+\frac{\mathbf{p}_{\lambda f}^2}{\alpha_{\lambda f}^2}\right)\right],
\end{split}
\label{eq:wfC}
\end{equation}
with only the spin $S=\frac{3}{2}$ component.
From the definition of Jacobi momentum, 
we can derive the two relations $\textit{\textbf{p}}_{Q}=-\textit{\textbf{p}}_{\lambda i}$ and $\textit{\textbf{p}}_{q}=-\textit{\textbf{p}}_{\lambda i}+\textit{\textbf{p}}_{f}$, which 
simplify the following calculation,
and set $\textit{\textbf{p}}_{f}$ along $z$-direction in the rest frame of the parent baryon.
Combining the explicit wave functions shown in Eq. (\ref{eq:wfC}) and delta function in Eq. (\ref{eq:matrixFFC}), we have
%
\begin{equation}
\begin{split}
\frac{|\vec{p}_f|}{2\sqrt{2}m_f}\bar{f}_1&=
\frac{1}{\pi^3}\frac{1}{(\alpha_{\rho i}\alpha_{\rho f}\alpha_{\lambda i}\alpha_{\lambda f})^{\frac{3}{2}}}N(\sigma_x)|_{(\frac{3}{2},\frac{1}{2})}\\
&\;\;\;\;\times\int(\frac{-\textit{\textbf{p}}_{\lambda i}+|\vec{p}_f|}{2m_q}+\frac{\textit{\textbf{p}}_{\lambda i}}{2m_Q})\exp\left[-\frac{1}{2}\left(\frac{\textit{\textbf{p}}_{\rho i}^2}{\alpha_{\rho i}^2}+\frac{\textit{\textbf{p}}_{\lambda i}^2}{\alpha_{\lambda i}^2}+\frac{\textit{\textbf{p}}_{\rho i}^2}{\alpha_{\rho f}^2}+\frac{(\textit{\textbf{p}}_{\lambda i}-\frac{2m}{m_f}|\vec{p}_f|)^2}{\alpha_{\lambda f}^2}\right)\right]d\textit{\textbf{p}}_{\rho i}d\textit{\textbf{p}}_{\lambda i}\;,
\end{split}
\end{equation}
with spectator quark mass $m$, leading to
\begin{equation}
\begin{split}
\bar{f}_1=\frac{2}{\sqrt{3}}\left(\frac{2\alpha_{\lambda i}\alpha_{\lambda f}}{\alpha_{\lambda i}^2+\alpha_{\lambda f}^2}\right)^{3/2}\left[1+\frac{2m\alpha_{\lambda f}^2}{m_q(\alpha_{\lambda i}^2+\alpha_{\lambda f}^2)}+\frac{2m\alpha_{\lambda i}^2}{m_Q(\alpha_{\lambda i}^2+\alpha_{\lambda f}^2)}\right]\;.
\end{split}
\label{eq:f1}
\end{equation}
Hence  the expression of $\bar{f}_1$ is derived.

\end{document}